\documentclass[iop]{emulateapj}
\usepackage{rotating}
\usepackage{apjfonts}
\usepackage{graphicx}
\usepackage{multirow}
\usepackage{comment}
\usepackage{natbib}
\usepackage{hyperref}
\usepackage{amsmath}
\usepackage{fontenc}
\usepackage{color}
\newcommand{\hbeta}{H{$\beta$}}
\newcommand{\halpha}{H{$\alpha$}}

\newcommand{\OIII}{[O{\sevenrm\,III}]}
\newcommand{\loiii}{$L_{\text{[O \textrm{\tiny III}]}}$}

\newcommand{\OIIIb}{[O{\sevenrm\,III}]\,$\lambda$5007}

\newcommand{\NII}{[N{\sevenrm\,II}]}

 \font\sevenrm=cmr7 scaled 1000

\def\chandra{{\it Chandra}}
\def\xmmnewton{{\it XMM-Newton}}
\def\swift{{\it Swift}}
\def\hst{{\it HST}}
\def\galfit{{\tt GALFIT}}
\def\sersic{S\'{e}rsic}
\newcommand{\mpsf}{$m_{\rm P}$}
\newcommand{\mser}{$m_{\rm S}$}
\newcommand{\reff}{$R_{\rm e}$}
\newcommand{\loiiiOBS}{$L^{\rm obs}_\text{[{\scriptsize O} \textrm{\tiny III}]}$}
\newcommand{\loiiiCOR}{$L^{\rm cor}_\text{[{\scriptsize O} \textrm{\tiny III}]}$}
\newcommand{\lbol}{$L_{\rm bol}$}
\newcommand{\REdd}{$\lambda_\mathrm{Edd}$}

\newcommand{\lum}{(\rm erg~s$^{-1}$)}
\newcommand{\rpy}{$r_{\mathrm{p}, Y}$}

\begin{document}

\title{Chandra X-ray and Hubble Space Telescope Imaging of
Optically Selected kiloparsec-Scale Binary Active Galactic
Nuclei II. Host Galaxy Morphology and AGN activity\altaffilmark{1}}

\shorttitle{kpc-Binary AGN II. Host Morphology and AGN Properties}

\shortauthors{SHANGGUAN ET AL.}
\author{Jinyi Shangguan\altaffilmark{2, 3}, Xin Liu\altaffilmark{4}, Luis C. Ho\altaffilmark{2, 3}, Yue Shen\altaffilmark{4}, Chien Y. Peng\altaffilmark{5}, \\
Jenny E. Greene\altaffilmark{6}, and Michael A. Strauss\altaffilmark{6}}

\altaffiltext{1}{Based, in part, on observations made with the NASA/ESA Hubble Space Telescope, obtained at the Space Telescope Science Institute, which is operated by the Association of Universities for Research in Astronomy, Inc., under NASA contract NAS 5-26555. These observations are associated with program number GO 12363.}

\altaffiltext{2}{Kavli Institute for Astronomy and Astrophysics, Peking University, Beijing 100871, China}

\altaffiltext{3}{Department of Astronomy, School of Physics, Peking University, Beijing 100871, China}

\altaffiltext{4}{Department of Astronomy, University of Illinois at Urbana- Champaign, Urbana, IL 61801, USA}

\altaffiltext{5}{Giant Magellan Telescope Organization, 251 South Lake Avenue, Suite 300 Pasadena, CA 91101, USA}

\altaffiltext{6}{Department of Astrophysical Sciences, Princeton University, Peyton Hall -- Ivy Lane, Princeton, NJ 08544, USA}

\begin{abstract}
Binary active galactic nuclei (AGNs) provide clues to how gas-rich mergers trigger and fuel AGNs and how supermassive black hole (SMBH) pairs evolve in a gas-rich environment.
While significant effort has been invested in their identification, the detailed properties of binary AGNs and their host galaxies are still poorly constrained.
In a companion paper, we examined the nature of ionizing sources in the double nuclei of four kpc-scale binary AGNs with redshifts between 0.1$\sim$0.2.
Here, we present their host galaxy morphology based on F336W ($U$-band) and F105W ($Y$-band) images taken by the Wide Field Camera 3 (WFC3) onboard the Hubble Space Telescope.
Our targets have double-peaked narrow emission lines and were confirmed to host binary AGNs with follow up observations.
We find that kpc-scale binary AGNs occur in galaxy mergers with diverse morphological types.
There are three major mergers with intermediate morphologies and a minor merger with a dominant disk component.
We estimate the masses of the SMBHs from their host bulge stellar masses and obtain Eddington ratios for each AGN.
Compared with a representative control sample drawn at the same redshift and stellar mass, the AGN luminosities and Eddington ratios of our binary AGNs are similar to those of single AGNs.
The $U-Y$ color maps indicate that clumpy star forming regions could significantly affect the X-ray detection of binary AGNs, e.g., the hardness ratio.
Considering the weak X-ray emission in AGNs triggered in merger systems, we suggest that samples of X-ray selected AGNs may be biased against gas-rich mergers.
\end{abstract}


\section{Introduction}\label{sec:intro}

In the $\Lambda$CDM cosmology of hierarchical structure formation (e.g., \citealt{White1978MNRAS}), the merger of galaxies is one of the major processes for galaxies to build up their stellar masses.
A merger of two galaxies that are massive enough to contain supermassive black holes \citep[SMBHs;][]{Kormendy1995ARAA, Kormendy2013ARAA} would trigger the accretion into SMBH(s) if tidal torques drive gas into the center of the gravitational potential, igniting the \textit{active galactic nuclei} (AGN; e.g., \citealt{Sanders1988ApJ}).
If the two SMBHs are triggered roughly at the same time, we would find a binary\footnote{In this context, "binary" does not require the SMBHs to be gravitationally bounded to each other.
For kpc-scale binary AGNs, the SMBHs still reside in the potential of their separate galactic nuclei.}
AGN.
Alternatively, binary AGN could also result from black holes in galaxy pairs that are simultaneously active by chance through stochastic accretion, that is not directly triggered by the merger \citep{Liu2012}.

Binary SMBHs expected to form in galaxy mergers \citep{Begelman1980Natur,yu02} are of great importance for the detection of gravitational waves (GWs; \citealt{Thorne1976ApJ, Thorne1987thyg.book,Centrella2010}) via pulsar timing arrays (PTAs; e.g., \citealt{Hobbs2010CQGra, Manchester2013PASA}) or space-borne GW observatories, e.g., eLISA/NGO \citep{Seoane2013arXiv}.
However, the evolution of binary SMBHs is still far from fully understood \citep{Dotti2012,Colpi2014SSRv,DEGN}.
\cite{Shannon2015arXiv} find their PTA data are inconsistent with current models of the GW background from the mergers of SMBHs \citep[e.g.,][]{McWilliams2014,Kulier2015}, which may indicate that the coalescence of the binary SMBHs are either stalled or accelerated.
The morphologies and kinematics of binary AGN host galaxies, especially, kpc-scale AGNs (referred to as binary AGNs, hereafter), are of unique importance for studies of binary-SMBH evolution, e.g., to compare with theoretical studies of binary SMBHs on multiple scales (e.g., \citealt{Mayer2007Sci, yu11,Fiacconi2013ApJ, Blecha2013a,Wassenhove2014MNRAS,Steinborn2015}).

The host galaxies of single AGNs and quasars have been studied extensively.
At low redshift ($z \lesssim 0.1$), AGN and quasar host galaxies tend to be bulge-dominated \citep[e.g.,][]{Ho1997ApJ, dunlop03,Ho2008ARAA} and are mainly in early-type galaxies with stellar masses rarely below $10^{10} \, \mathrm{M_\odot}$ \citep{Kauffmann2003MNRAS}.
There is also growing evidence that AGN host galaxies at intermediate redshifts (up to $z \sim 3$) also display bulge-dominated or bulge+disk morphologies \citep{Grogin2005ApJ, Gabor2009ApJ, Povic2012AA, Kocevski2012ApJ, Rosario2015AA, Bruce2015arXiv}.
It is possible that violent processes, e.g., major mergers \citep{Hopkins2005ApJ} or violent disk instabilities \citep{Bournaud2011ApJ, Bournaud2015arXiv}, could drive gas into nuclei to fuel SMBHs and build up the galactic bulges.
Gas-rich major mergers are often suggested to trigger the most luminous AGNs and quasars \citep{Sanders1988ApJ, Hernquist1989Natur, Kauffmann2000MNRAS, Hopkins2008ApJS}. While studies of quasar host galaxies have found a high fraction of mergers \citep[e.g.,][]{bahcall97,kirhakos99} or a dense local environment \citep[e.g.,][]{serber06}, more recent morphological studies have brought the merger scenario into question for moderate luminosity AGNs and quasars \citep{Grogin2005ApJ, Gabor2009ApJ, Cisternas2011ApJ, Schawinski2011ApJ, Kocevski2012ApJ, Villforth2014MNRAS, Bruce2015arXiv,Mechtley2015}.
They find the same fraction of mergers in the host galaxies of AGNs as normal galaxy control samples.
Meanwhile, other studies find a higher fraction of mergers in AGN hosts than in control samples of inactive galaxies \citep{Ellison2008AJ, Ellison2011MNRAS, Liu2012,Ellison2013MNRAS, Koss2010ApJ, Koss2012ApJ, Almeida2011MNRAS, Silverman2011ApJ,Chiaberge2015ApJ,Rosario2015AA}.  Binary AGNs provide a unique sample to study AGN activity in ongoing mergers.

The search for binary AGNs is still challenging \citep{popovic11}.
In addition to serendipitous discoveries (e.g., \citealt{Moran1992AJ, Junkkarinen2001ApJ, Komossa2003ApJL, Hudson2006AA, Bianchi2008MNRAS, Koss2011ApJL, fabbiano11,Shields2012,Huang2014MNRAS}), systematic searches have been conducted at many wavelengths (e.g., \citealt{green10,Liu2011ApJ, Koss2012ApJ, Fu2015ApJ}). Many searches use double-peaked narrow emission lines as an indicator to select candidate binary AGNs \citep{Comerford2009ApJ, Wang2009ApJL, Smith2010ApJ, Liu2010ApJ, Liu2011ApJ, Ge2012ApJS, Comerford2013ApJ} using the Sloan Digital Sky Survey \citep[SDSS;][]{York2000AJ} galaxy sample.
If the two nuclei of a binary AGN are within $3''$ (5.5 kpc at $z\sim0.1$), the spectra of the two AGNs fall within a single fibre.
The integrated spectrum then displays double-peaked emission lines, originating from two narrow line regions (NLRs) with different light-of-sight velocities.
However, the double peaked emission lines could also originate from various other processes, such as biconical outflows, gas rotations in NLRs and jet-cloud interactions \citep[e.g.,][]{crenshaw09,xu09,fischer11, Shen2011ApJ, Fu2011ApJa}.
Thus, careful diagnoses and/or follow-up observations are necessary to confirm the existence of double AGN activity \citep{Liu2010ApJL,Comerford2011ApJ, Fu2011ApJa, Fu2011ApJb, Fu2012ApJ, Shen2011ApJ, Liu2013ApJ, Comerford2015ApJ, McGurk2015ApJ, Muller-Sanchez2015arXiv}.

While significant effort has been invested in identification on binary AGNs and their host galaxies, there are few studies in their detailed characterization.
Using adaptive optics at the Keck II telescope \citep{Wizinowich2000PASP, Wizinowich2000SPIE}, \cite{Max2007Sci} obtained high-resolution near-infrared images to study the host galaxies of NGC 6240, a prototypical binary AGN. 
Combined with radio and X-ray observations, they find that each of the two central BHs is at the center of a rotating stellar disk, surrounded by a cloud of young star clusters.
\cite{Villforth2015AJ} took deep images and long-slit spectra for four targets with double-peaked narrow emission lines.
They found that 3 targets are undergoing mergers, with unambiguous tidal features and powerful outflows.
The remaining one, SDSS J1715+6008, confirmed to be a binary AGN by \cite{Comerford2011ApJ} with X-ray detection of both nuclei albeit with very low counts, display so well relaxed morphology, indicating that the merger, if any, happened $\gtrsim 1$ Gyr ago.
Additionally, the misalignment of gas and stellar kinematics implies bipolar outflows or a counter-rotating gas disk in this galaxy.
Their work highlights the significance of binary AGNs as a laboratory to study the complicated gas and stellar dynamics in mergers.

In this paper, we aim to understand the uniqueness of binary AGNs and their host galaxies for these ongoing mergers.
What are the morphological features observed in binary AGN host galaxies?
How do mergers affect AGN properties?
These are the main goals of our present work.
More specifically, we will measure the masses of the host galaxies, the stellar bulges and the SMBHs.
We can then examine the Eddington ratios, using estimates on the AGN bolometric luminosities, and put them into context by comparing with control samples or ordinary AGNs.

The rest of the paper is organized as follows. We describe our sample in Section \ref{sec:sample} and data reduction in Section \ref{sec:hst}.
We measure the photometric magnitudes in both bands (\S \ref{subsec:mag}), generate color maps (\S \ref{subsec:cmp}) and fit the $Y$-band surface brightness profiles of the host galaxies (\S \ref{subsec:galfit}).
In Section \ref{sec:result}, we investigate the morphologies and color maps of the host galaxies (\S \ref{subsec:mph}), and measure the stellar masses of the galaxies and the bulges (\S \ref{subsec:mass}).
We estimate the SMBH masses, bolometric luminosities and Eddington ratios of the AGNs and compare their properties with those of single AGNs (\S \ref{subsec:bol}).
We discuss the relations between the color maps and X-ray emission in \S \ref{subsec:x+c}, illustrating how the X-ray emission is affected by nuclear star formation.
Based on the putative feature of X-ray weakness in binary AGNs, we suggest that previous studies of AGN host galaxies using X-ray selected samples might be biased against merger systems (\S \ref{subsec:bias}).

Throughout this paper, we assume a concordance cosmology with $\Omega_m = 0.3$, $\Omega_{\Lambda} = 0.7$, and $H_{0}=70$ km s$^{-1}$ Mpc$^{-1}$, and use the AB magnitude system \citep{Oke1974ApJS}.

\section{The Sample}\label{sec:sample}

Here and in a companion paper \citep[][paper I]{Liu2013ApJ}, we present Hubble Space Telescope (HST) Wide Field Camera 3 (WFC3) $U$- and $Y$-band images and \chandra\ X-ray Observatory \citep[CXO;][]{Weisskopf1996SPIE} Advanced CCD Imaging Spectrometer \citep[ACIS;][]{Garmire2003SPIE} 0.5--10 keV X-ray images of four optically selected kpc-scale binary AGNs \citep[Table \ref{tab:obs};][]{Liu2010ApJL}.
The binary AGN candidates were identified from the double-peaked narrow-line AGN sample of \citet{Liu2010ApJ} selected from the SDSS Data Release Seven \citep[DR7;][]{Abazajian2009ApJS}.
In each system, our NIR images show tidal features and double stellar components with a projected separation of several kpc, while our optical slit spectra show to Seyfert 2 nuclei spatially coincident with the stellar components, with line-of-sight velocity offsets of a few hundred km s$^{-1}$ \citep{Liu2010ApJL}.
While our ground-based NIR imaging and spatially resolved optical spectroscopy strongly suggest that these galaxy mergers host binary AGNs, the case was not watertight.
To further clarify the ambiguities associated with the optical classification, in paper I, we examined the nature of the ionizing sources in the double nuclei combining \chandra\ X-ray imaging spectroscopy and \hst\ $U$-band imaging.
SDSS J1108+0659 and SDSS J1146+5110 (J1108 and J1146) are confirmed to be binary AGNs.
For the other two targets, SDSS J1131-0204 and SDSS J1332+0606 (J1131 and J1332), the current data are still consistent with the binary AGN scenario, but the possibility of only one AGN ionizing both components in the merger cannot be ruled out.
Combining with previous optical spectroscopy, we found tentative evidence for a systematically smaller hard-X-ray-to-\OIII\ luminosity ratio (see also \citealt{Comerford2015ApJ}) and/or higher Compton-thick fraction in optically selected kpc-scale binary AGNs than in optically selected single Type 2 Seyferts.

\begin{figure*}
\begin{center}
\includegraphics[width=0.95\textwidth]{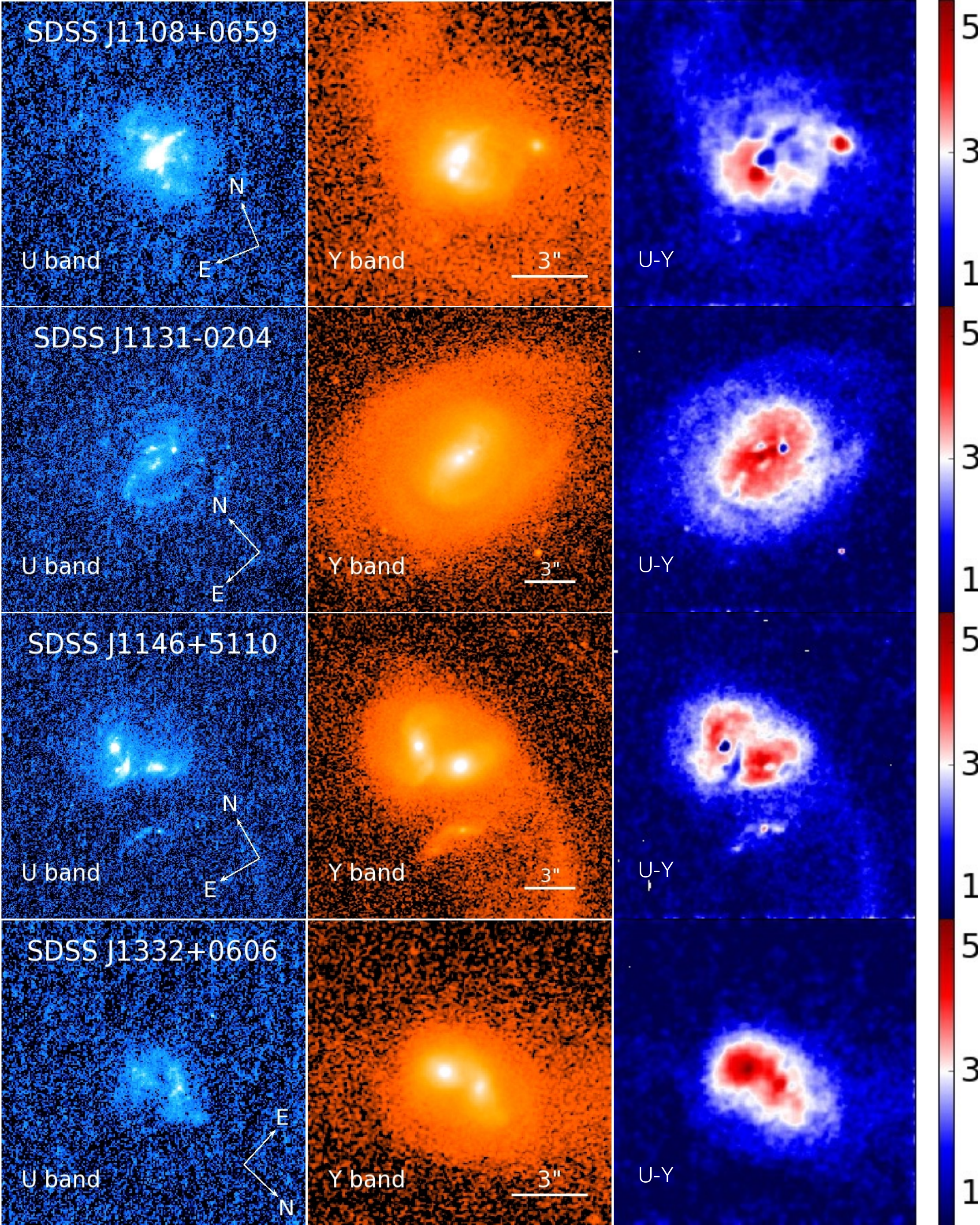}
\caption{HST/WFC3 F336W ($U$-band, {\it left column}), F105W ($Y$-band, {\it middle column}), and $U-Y$ color images of the four kpc-scale binary AGNs.
Images in each row have the same FOV and orientation.
The color code is $m_U - m_Y$.
}
\label{fig:cmp}
\end{center}
\end{figure*}

\section{Hubble Space Telescope Imaging}\label{sec:hst}

The four optically selected kpc-scale binary AGNs were observed using the WFC3 onboard the \hst\ in Cycle 18 under program GO 12363, between 2011 March and June (with observing dates listed in Table \ref{tab:obs}).
Each target was imaged in the UVIS/F336W \citep[$U$ band, with pivot $\lambda_{p}=335.5$ nm and width of 51.1 nm;][]{Dressel2012WFC3} and IR/F105W (wide $Y$ band, with pivot $\lambda_{p}=1055.2$ nm and width of 265.0 nm) filters within a single {\it HST} orbit, with total net exposure times being $\sim 2100 \, \mathrm{s}$ in the F336W filter and $239 \, \mathrm{s}$ in the F105W filter, respectively.
The pixel size of the WFC3 UVIS CCD (IR detector) is $0 \farcs 039$ ($0 \farcs 13$).
The F336W (F105W) observations were dithered at three (four) positions to properly sample the point spread functions (PSFs) and to reject cosmic rays and bad pixels.
A 1k$\times$1k (512$\times$512) sub-array was employed for the F336W (F105W) imaging, yielding a field of view (FOV) of $40''\times37''$ ($72''\times64''$), adequate to cover our targets and the nearby field for background subtraction.

We reduced the WFC3 data following standard procedures using the \textsf{calwf3} and \textsf{AstroDrizzle} in PyRAF.
After \textsf{calwf3} reduction, both $U$- and $Y$-band images were processed with \textsf{AstroDrizzle} to correct for geometric distortion and pixel area effects.
Dithered frames were combined rejecting cosmic rays and hot pixels. The pixel scale of the final image product is chosen to be $0 \farcs 06$, so that the $Y$-band images are Nyquist sampled.
The reference pixels and orientation of $U$-band images are also matched with the $Y$-band images for generating the color maps.

\subsection{Absolute and Relative Astrometry of WFC3 Images}

Because \textsf{MultiDrizzle} relies on the measured and catalog positions of guide stars for absolute astrometric calibration, the absolute astrometric accuracy of a WFC3 image processed by \textsf{MultiDrizzle} is limited by the positional uncertainty of guide stars ($\gtrsim 0\farcs2$) and the calibration uncertainty of the fine guidance sensor to the instrument aperture ($\sim0\farcs015$).
The relative astrometric accuracy of WFC3 images is primarily limited by the uncertainty from the geometric distortion correction of the camera.
Typical relative astrometric accuracy is $0\farcs004$ for the UVIS and $0\farcs01$ for the IR images.

We have checked that the astrometry of our $Y$-band images agrees with the SDSS within $0\farcs4$, based on positions of the matched sources within the FOV.
To achieve better absolute astrometric accuracy, we registered the $Y$-band images with the SDSS astrometry \citep{pier03}.
Reference objects were first selected from the FOV with known SDSS astrometry.
We rejected objects with high proper motion and/or extended surface brightness profiles as outliers in the WCS fitting using the \textsf{msctpeak} task in IRAF.
The typical statistical uncertainty from the fitting was $\sim0\farcs085$.

Figure \ref{fig:cmp} shows the calibrated $Y$- and $U$-band images of our targets.
We estimate the absolute astrometric uncertainty of the registered $Y$-band images as $\sim 0\farcs15$.
This includes $\sim0\farcs055$ which is the typical uncertainty of SDSS astrometry (combining the statistical error per coordinate $\sim0\farcs045$ and the systematic error $0\farcs03$ as well as possible additional systematic error $0\farcs01$ due to proper motion; \citealt{pier03}) and $\sim0\farcs085$ which is the statistical uncertainty estimated from the standard deviation of the WCS fits.
Paper I lists the coordinates of each of the double nuclei and the inferred nuclear separations for each target.
These $Y$-band nuclear separations agree with those measured from our ground-based NIR imaging within uncertainties.
The $Y$-band images improve upon the accuracy of the expected locations of the double nuclei.
Similarly we have registered the $U$-band images with the SDSS astrometry, using reference objects detected in the FOV.
The estimated absolute astrometric uncertainty of the registered $U$-band images is $\sim 0\farcs20$, which is larger than that for the $Y$-band images, since fewer reference objects were detected.

\begin{table}
\caption{The Four kpc-scale Binary AGN Targets}
\begin{center}
\begin{tabular}{ c c c c c c c c c}
  \hline
  \hline
Target Name & Redshift & $\Delta \theta_\mathrm{NIR}$ & $\Delta \theta_\mathrm{Y}$ & \rpy  \\
  & $z_c$  & ($''$) & ($''$) & (kpc)  \\
(1) & (2) & (3) & (4) & (5)     \\ \hline
SDSS J110851.04+065901.4   & 0.1816 & 0.5  & 0.81 & 2.47  \\
SDSS J113126.08$-$020459.2 & 0.1463 & 0.6  & 0.76 & 1.96  \\
SDSS J114642.47+511029.6   & 0.1300 & 2.7  & 2.73 & 6.32  \\
SDSS J133226.34+060627.4   & 0.2070 & 1.5  & 1.52 & 5.16  \\ \hline
\end{tabular}
\end{center}
\tablecomments{
Col. 1: SDSS names with J2000 coordinates given in the form of "hhmmss.ss$\pm$ddmmss.s".
Col. 2: systemic redshift measured from stellar continuum absorption features in the SDSS fiber spectra.
Col. 3: angular separation between the double nuclei measured from the NIR images as reported in \citet{Liu2010ApJL}.
Cols. 4 \& 5: separation between the double nuclei measured from HST $Y$-band images through \galfit\ fitting (see \S \ref{subsec:galfit}).
}
\label{tab:obs}
\end{table}

\subsection{Aperture Photometry of the Host Galaxies}\label{subsec:mag}

We measure the magnitudes of our targets with contaminating sources masked.
We use \texttt{SEP} \citep{Barbary2015SEP} to generate masks for all sources in the images.
Using \galfit\ modeling (\citealt{Peng2002AJ, Peng2010AJ}; see \S \ref{subsec:galfit} for detail description), we find that the overlapping light from the neighboring objects are negligible, if there is any.
For $U$-band images (\textit{first column} of Figure \ref{fig:cmp}), we employ the same masks obtained for $Y$ band, since all the nearby galaxies appearing in $Y$ band are potential contamination sources in the $U$ band, even though they are much less visible there.
The aperture radii are chosen to be $18''$ for J1108 and J1146 and $12''$ for J1131 and J1332 in both bands.
By studying the growth curves, we find the apertures are large enough to enclose more than 95\% of the total fluxes in $Y$ band, so the flux uncertainties due to aperture size are less than 0.06 mag.
The measured $Y$- and $U$-band magnitudes are listed in Table \ref{tab:mag}, with uncertainties measured from sky variations (see discussion below).

To determine the background sky, we randomly sample $50 \times 50$ squares of pixels in each masked image.
The mean of the resulting counts (median values of pixels in the squares) is subtracted from the image, and the standard deviation is used to estimate the uncertainties of the measured magnitudes.
The host galaxies of binary AGNs are always very extended in both bands, as expected for relatively early-stage mergers.
For U-band images, in particular, we need to use large masks, more than encompassing the apparent size of the galaxies, to determine the true width of the sky distribution.


\subsection{$U$-band Images and $U-Y$ Color Maps}\label{subsec:cmp}

The $U$-band images of the galaxies (Figure \ref{fig:cmp} {\it left column}) show clumpy structures spreading over most of the galaxies, indicating clumpy star-forming regions which are common in gas-rich mergers.
Figure \ref{fig:cmp} ({\it right column}) shows the $U-Y$ color maps of our four targets.
We convolve the $U$- and $Y$-band images ({\it left} and {\it middle} columns) with Gaussian kernels with different sizes to match the PSFs in the two bands to be FWHM $\approx \, 0\farcs14$ arcsec.
The pixels below 1 $\sigma$ of sky Poisson noise in either band are replaced by the interpolated values using the nearby valid pixels and the Gaussian kernel used in the convolution as the interpolation function\footnote{This is achieved by using the {\tt convolve} function in the Astropy package \citep{Astropy2013AA}.}.
Since the center and orientation of the images in two bands are matched with each target, we generate the color map pixel by pixel, with $U-Y = -2.5 \, \mathrm{log_{10}} \frac{f_U}{f_Y} + m_{0, U} - m_{0, Y}$, where $f_U$ and $f_Y$ are the fluxes at corresponding pixels in $U$ and $Y$ bands and $m_{0, U}=24.64$ and $m_{0, Y}=26.27$ are the zero points of the two bands\footnote{\url{http://www.stsci.edu/hst/wfc3/phot_zp_lbn}}.

\begin{table*}
\caption{Photometric Parameters, Host Galaxy Stellar Masses, and Bulge-to-total Ratios}
\begin{center}
\begin{tabular}{c c c c c c c c}
  \hline
  \hline
  Target Name     & $m_Y$           & $m_U$           & $M_z$  & $M_u$  & $u-z$          & log $M_*$ & B/T  \\
      & (mag)           & (mag)            & (mag)   & (mag)   & (mag)           & ($M_{\odot}$)  &   \\
      (1)         &    (2)   &  (3)            &  (4)            &   (5)  &   (6)  &  (7)           &  (8)   \\ \hline
 J1108   & 16.24$\pm 0.05$ & 18.48$\pm 0.26$ & $-$23.12 & $-$21.37 & 1.76$\pm 0.26$ & 11.14 & 0.55 \\
 J1131 & 15.81$\pm 0.01$ & 18.94$\pm 0.09$ & $-$23.04 & $-$20.39 & 2.65$\pm 0.10$ & 11.24 & 0.12 \\
 J1146   & 15.89$\pm 0.03$ & 18.40$\pm 0.14$ & $-$22.68 & $-$20.65 & 2.02$\pm 0.14$ & 11.00 & 0.41 \\
J1332   & 16.82$\pm 0.04$ & 19.29$\pm 0.16$ & $-$22.86 & $-$20.88 & 1.98$\pm 0.17$ & 11.07 & 0.69 \\ \hline
\end{tabular}
\end{center}
\tablecomments{
Col. 2 \& 3: $Y$- and $U$-band apparent magnitudes (see \S \ref{subsec:mag}). The uncertainties are estimated from 1 $\sigma$ variation of the sky background;
Col. 4 \& 5: SDSS $z$- and $u$-band absolute magnitudes of the targets, transformed from $m_Y$ and $m_U$ respectively, assuming a flat local spectrum ($f_{\lambda} \sim \mathrm{constant}$) around the relevant wavelengths;
Col. 6: Color calculated from $u$- and $z$-band magnitudes;
Col. 7: Total galaxy stellar mass of the targets estimated from $z$- and $u$-band magnitudes (see \S \ref{subsec:mass} for details,).
Col. 8: Bulge-to-total stellar mass ratios. The bulge stellar mass is obtained by adding up the two bulge components in \galfit\ decomposition results for each galaxy (Table \ref{tab:bol}).
}
\label{tab:mag}
\end{table*}

\subsection{Surface Brightness Profile Fitting of $Y$-band Images}\label{subsec:galfit}

The $Y$-band images probe old stellar populations, allowing us to explore detailed host galaxy morphology and low surface brightness tidal features indicative of mergers.
We use \galfit\, a two-dimensional fitting algorithm, to model the multiple structural components in the $Y$-band images of our galaxy-merger targets.
We aim to decompose the double nuclei and the associated bulge components, if any, and to measure the low surface brightness tidal features contained in the high quality and high sensitivity \hst\ images.

\begin{figure*}[!htpb]
\begin{center}
\includegraphics[width=0.95\textwidth]{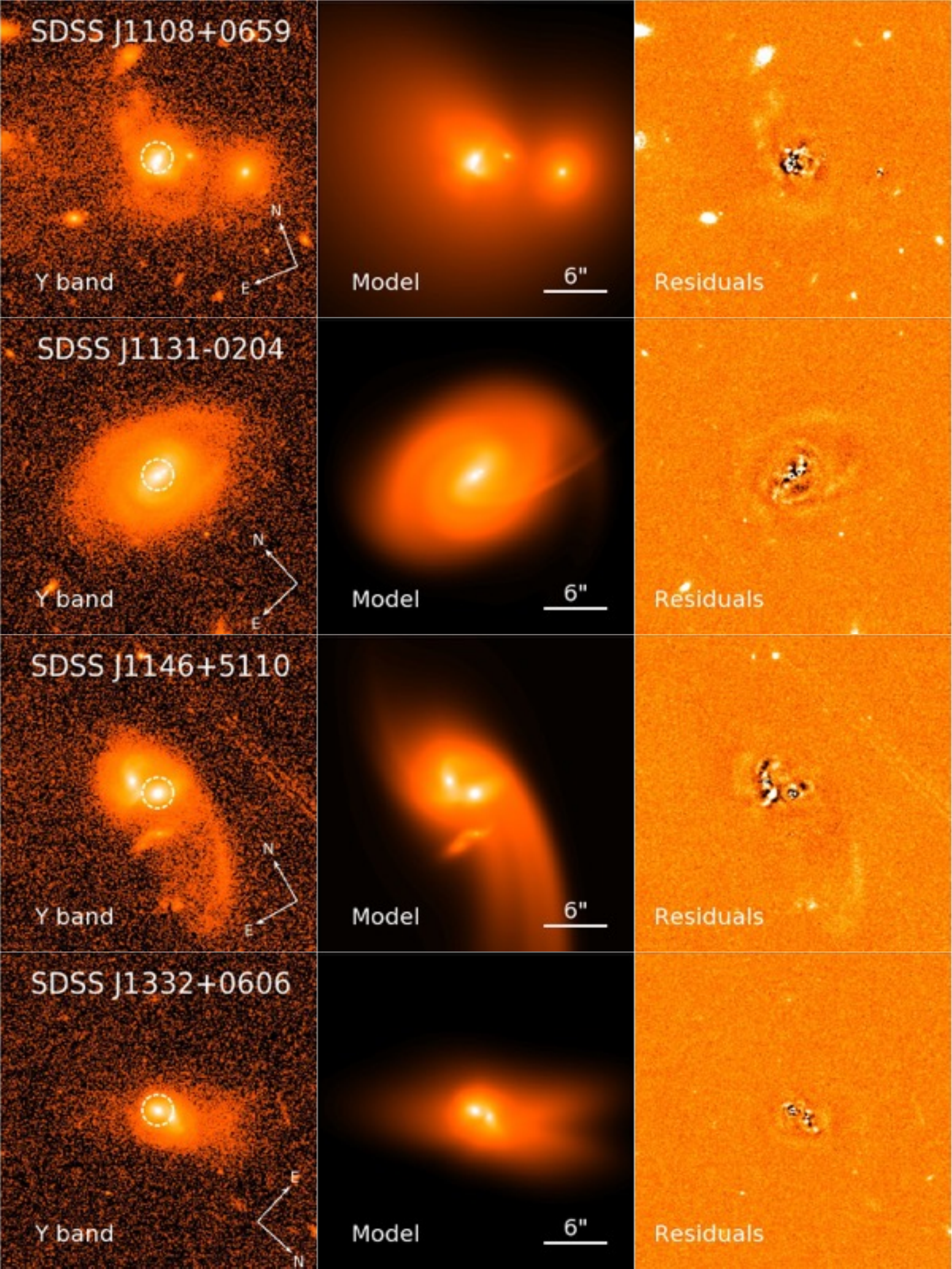}
\caption{
Model fitting of the \hst\ $Y$-band images of our four binary AGN targets.
The field shown is larger than in Figure. \ref{fig:cmp}.
{\it Left column:} The original $Y$-band images.
The white dashed circles indicate the spatial coverage of the SDSS fibers ($3''$ in diameter). {\it Middle column:} Our best-fit models from \galfit\ analysis.
The pixel brightness scale is logarithmic in the first two columns.
The residuals are shown in the {\it right column}, where the pixel brightness scale is linear.
We simultaneously fit the close companions of J1108 and J1146 because they are partially blended with our targets.
}
\label{fig:galfit}
\end{center}
\end{figure*}

\galfit\ is well suited for our goals.  The spiral and Fourier modes of \galfit\ are powerful to fit the irregular shapes of the merging systems.
We use stars within the FOV to model the PSFs.
We mask out the relatively isolated companions and fit projected nearby companions simultaneously in order to fit our targets more accurately.
A constant sky background is employed for every target.
For baseline models we first adopt one \sersic\ component \citep{Sersic1963BAAA, Sersic1968adga} to fit each nucleus and another to fit the extended stellar envelope.
We use PSFs to measure any unresolved nuclei, which turn out to be necessary in most cases, although they are always very faint.
These unresolved components could be related to nuclear starbursts \citep{Bruce2015arXiv}, but they are in general too faint to significantly affect our results (e.g., bulge masses).
The results are shown in Figure \ref{fig:galfit} and summarized in Table \ref{tab:galfit}.

For J1108 and J1332, we regard the \sersic\ component for each nucleus as the bulge (Table \ref{tab:galfit}).
The J1131E nucleus and the J1146SW nucleus consist two \sersic\ components in their model.
We choose the inner one with the higher \sersic\ indices as the bulges.
We fit J1108 as well as two companions to the south west simultaneously.
For the minor merger system (see discussion below), J1131, the weaker nucleus is so faint that it is best fit by a PSF.
Meanwhile, the barred bulge of the brighter nucleus requires two \sersic\ components.
For J1146, the system is so complicated that we need another \sersic\ component to fit the southern nucleus.
There is another small disturbed galaxy to the south east which is so blended with J1146 that we have to fit them together.
We still see some patterns in the residual images (Figure \ref{fig:galfit} \textit{right column}), reflecting the fact that our targets are highly disturbed.
The residuals are generally orders of magnitude fainter than the overall values for each pixel ($\lesssim 1\%$ on average), indicating that our fits reproduce the main features of the galaxies.
The nominal fitting errors are usually very small ($\lesssim 1\%$), but the real uncertainties are dominated by systematic errors due to decomposition and model uncertainty.
However, systematic uncertainties from fitting highly disturbed merging systems are difficult to quantify doing so is beyond the scope of this paper.
The separations of the two nuclei (\rpy\ in Table \ref{tab:obs}) are estimated from the separations of the PSFs in each targets.
The closest pair (J1131) is separated by 2.0 kpc, while the farthest one (J1146) are 6.3 kpc.

\section{Results}\label{sec:result}

\subsection{Host Galaxy Morphology}\label{subsec:mph}

The \hst\ $Y$-band images reveal the detailed morphologies of the host galaxies (see Figure \ref{fig:cmp} {\it middle column} and Figure \ref{fig:galfit} {\it left column} with larger FOV).
It is apparent that all the host galaxies are mergers with double stellar nuclei.
J1108, J1146 and J1332 are major mergers (see discussion below) where the two stellar nuclei are comparably bright.
The \sersic\ indices suggest that the host galaxies are intermediate types ($n \approx 1-2.5$), typical of single AGNs at these redshifts \citep{Kauffmann2003MNRAS}.
J1131 is a minor merger which displays a clear dominant disk structure with bars and spiral arms.
While disk structures are generally expected to be disrupted in major mergers, there is still the possibility that the disk reforms or survives \citep{Robertson2006ApJ, Hopkins2009ApJ}.

Tidal features are prominent in the $Y$-band images in all cases.
J1108 displays an extended low-surface brightness component to the east of the galaxy.
It is surrounded by several projected close companions that could be physically related.
J1131 shows features of perturbation in the spiral disk.
A bar structure is observed along the line of the two nuclei.
J1146 has an extended tidal tail and the bulges are elongated and twisted.
J1332 also displays extended tidal features and its bulges appear to be distorted.

By selection all of our targets are at relatively early merger stages where the double stellar nuclei are still resolvable, although they are also advanced enough that the bulges are difficult to separate unambiguously, especially for J1108 and J1131.
There is considerable overlap of the two bulges in J1146 and J1332 as well.
Therefore, \galfit\ fitting is necessary to decompose and measure each stellar component.

There are clear gradients in the color maps (Figure \ref{fig:cmp}).
The galaxies are redder towards the inner regions, blue clumps are apparent in and around the inner regions.
These clumps are most prominent in the west of J1108.
For J1146, one of the inner spiral arms of the north-east nucleus is blue.
The blue clumps highlight star forming regions that are not heavily obscured by dust.
In J1108, J1131 and J1146, there is one red and one blue nucleus residing in each galaxy.
The color of the nuclei may be affected by three factors: AGN activity, nuclear star formation and dust attenuation.
Since our targets are all type 2 AGNs \citep{Liu2010ApJL}, we do not expect the AGNs to emit strong $U$-band radiation (see Paper I).
Thus, the excess of $U$-band photons in the blue nuclei likely comes from local star formation activity.
The association between the soft X-ray emission and optical colors of the nuclei also support this explanation (see \S \ref{subsec:x+c} for details).
However, it is possible that the scattered AGN light generates the blue clumps \citep{Zakamska2006AJ}.

\subsection{Host Galaxy Stellar Mass}\label{subsec:mass}

We estimate stellar masses for the host galaxies of our binary AGN targets based on the $Y$-band magnitudes and $U-Y$ colors.
We apply k-corrections to convert $m_Y$ and $m_U$ into SDSS $z$- and $u$-band magnitudes\footnote{$m_z = m_Y + 0.36$ and $m_u = m_U - 0.13$} at rest frame (Table \ref{tab:mag} Col. 3-6), assuming a flat local spectrum ($f_\lambda \sim \mathrm{constant}$) over the relevant spectral range.
Since the corresponding bands are very close, the error induced from the k-correction is much smaller than the scatter of the stellar mass-to-light ratio described below. We use $z$-band absolute magnitude, $M_z$, and color $u-z$ to estimate the stellar masses based on the stellar $M/L$ ratio empirical relation provided by \cite{Bell2003ApJS},
\begin{equation}\label{eq:mass}
\mathrm{log_{10}}\left(\frac{M_*}{M_\odot}\right) = -0.4 \, (M_z - M_{z, \odot})
- 0.179 + 0.151 \, (u-z),
\end{equation}
where $M_*$ is the galaxy stellar mass in solar units and $M_{z, \odot}=4.51$ is the absolute magnitude of the Sun in the $z$ band \citep{Blanton2007AJ}.
According to the empirical calibration by \cite{Bell2003ApJS}, systematic uncertainties from galaxy age, dust and bursts of star formation are typically 0.1 dex for optical M/L ratios and are larger for galaxies with bluer optical colors.
The $M_*$ estimates are generally $\sim 10^{11} M_\odot$, consistent with measurements based on SDSS data (\citealt{Liu2010ApJL}).

We then employ the bulge stellar masses to estimate the mass ratios of the merging components and the bulge-to-total ratios (B/Ts) of the descendent galaxies in the mergers.
However, individual bulge masses are difficult to measure since they overlap with each other.
We can only estimate the bulge masses based on our \galfit\ decomposition of the $Y$-band images.
The bulge $z$-band absolute magnitudes (k-corrected from $Y$ band in the same way as discussed above) and stellar masses are listed in Table \ref{tab:bol}.
The J1131W nucleus is best fit by a PSF, thus we could not measure its bulge magnitude.
Since the bulge magnitudes in the $U$ band are hard to measure by \galfit\ fitting, we adopt the overall color of the galaxy (Table \ref{tab:mag}) to estimate the bulge mass.
While the colors of the bulges vary across different parts within the same galaxy (see \S \ref{subsec:cmp}), our approximation of using the overall colors as a surrogate for the bulge colors generally do not affect our main results\footnote{According to the $U-Y$ color maps, the bulge colors deviate $\lesssim 1.0 \, \mathrm{mag}$ from the galaxy colors, giving a $\lesssim 0.15$ dex variation in stellar mass (Eqn. \ref{eq:mass}). }.

From the mass ratios of the bulges we examine the merger types of the host galaxies of our binary AGN targets.
The bulge mass ratios are roughly 1:1 for J1108, J1146 and J1332; in J1131, the minor bulge is too small to measure.
This is not surprising because our selection method with double-peaked narrow emission lines preferentially picks out equally strong AGNs \citep{Liu2010ApJ}.
For each host galaxy, we add up the bulge masses to obtain the overall B/Ts (Table \ref{tab:mag}).
The B/Ts for our major mergers are all $\gtrsim 0.4$.
Since these are only lower limits for the descendent galaxies, the three major mergers are likely to become bulge dominant after the merger.

Our results have two main caveats.
First, since the mergers are on-going, the interaction of the two galaxies will continue heating stars in the disk, contributing to the growth of the bulge.
Thus, the observed bulges could be larger than the original bulges (which presumably followed the bulge-SMBH mass relation), while the the sum of the two bulges may be smaller than the bulge of the fully relaxed merged galaxy.
Second, our measurements are based on the \galfit\ fitting results.
The bulge mass may only account for a fraction of one \sersic\ component.
Nevertheless, the correlation of Eddington ratios and \OIIIb\ (hereafter, \OIII]) luminosity of our targets, which we discuss in Section \ref{subsec:bol}, may indicate that the bulge mass estimates are indeed related to the SMBH masses (see discussion in \S \ref{subsec:bol}).

\begin{table}
\caption{HST $Y$-band Photometric Decomposition Results from GALFIT Analysis}
\begin{center}
\begin{tabular}{ c  c  c  c  c  c  c  c  c }
  \hline
  \hline
  Target Name                    & \multicolumn{2}{c}{J1108} & \multicolumn{2}{c}{J1131} &
                            \multicolumn{2}{c}{J1146} & \multicolumn{2}{c}{J1332} \\\hline
\multirow{4}{*}{Nucleus 1}           & \mpsf & 21.2 & \mpsf & 21.1 & \mpsf & 19.2 & \mpsf & 20.5 \\\cline{2-9}
                                     & \mser & 17.4 &       &      & \mser & 17.5 & \mser & 17.8 \\
                                     & \reff &  1.8 &       &      & \reff &  3.2 & \reff & 2.13 \\
                                     & $n$   &  1.5 &       &      & $n$   &  2.3 & $n$   &  1.4 \\\hline
\multirow{7}{*}{Nucleus 2}           & \mpsf & 20.5 & \mpsf & 21.1 & \mpsf & 20.5 & \mpsf & 22.5 \\\cline{2-9}
                                     & \mser & 18.0 & \mser & 18.1 & \mser & 17.8 & \mser & 18.2 \\
                                     & \reff &  2.8 & \reff &  1.8 & \reff &  0.7 & \reff &  6.6 \\
                                     & $n$   &  1.0 & $n$   &  0.7 & $n$   &  1.1 & $n$   &  1.7 \\\cline{2-9}
                                     &       &      & \mser & 17.0 & \mser & 17.9 &       &      \\
                                     &       &      & \reff &  7.6 & \reff &  3.6 &       &      \\
                                     &       &      & $n$   &  0.5 & $n$   &  0.3 &       &      \\\hline
\multirow{3}{*}{Extended Envelope}   & \mser & 17.0 & \mser & 16.5 & \mser & 17.0 & \mser & 18.2 \\
                                     & \reff & 18.9 & \reff & 13.0 & \reff & 14.1 & \reff & 17.5 \\
                                     & $n$   &  1.5 & $n$   &  0.5 & $n$   &  1.2 & $n$   &  0.7 \\\hline
\end{tabular}
\end{center}
\tablecomments{
Nucleus 1 denotes J1108NW, J1131W, J1146NE and J1332SW, respectively for each target, whereas Nucleus 2 denotes J1108SE, J1131E, J1146SW and J1332NE.
Our baseline models adopt one PSF and one \sersic\ component to fit each nuclei and another \sersic\ component to fit the extended region.
But for J1131, which is a minor merger, Nucleus 1 is best fit by one PSF and Nucleus 2 is best fit by two \sersic\ components.
For J1146, Nucleus 2 is best fit by two \sersic\ components.
The integrated magnitudes of the PSF and \sersic\ components are \mpsf\ and \mser , respectively.
\reff\ (in units of kpc) and $n$ are the effective radius and \sersic\ index of each \sersic\ component.
The random errors from the fit are generally small ($\lesssim 1\%$) but the uncertainties are dominated by systematic errors due to priors about the model profiles and components. 
See text for more details.
}
\label{tab:galfit}
\end{table}%

\subsection{Black Hole Mass and Eddington Ratio}\label{subsec:bol}

We estimate the masses of the SMBHs (Table \ref{tab:bol}) using the $M_\mathrm{BH}$--$M_\mathrm{bulge}$ relation \citep{Kormendy2013ARAA},
\begin{equation}
\frac{M_\mathrm{BH}}{10^9 \, M_\odot} = 0.49
  \left(\frac{M_\mathrm{bulge}}{10^{11} \, M_\odot}\right)^{1.17},
\end{equation}
with an intrinsic scatter of $0.28 \, \mathrm{dex}$.  The uncertainty of $M_\mathrm{BH}$ is $0.3 \, \mathrm{dex}$, including the bulge-mass uncertainty ($\sim 0.1 \, \mathrm{dex}$; \S \ref{subsec:mass}) and the intrinsic scatter of the $M_\mathrm{BH}$--$M_\mathrm{bulge}$ relation. The calibration by \cite{Kormendy2013ARAA} yields black hole masses several times larger than previous calibrations, which were attributed by the authors in part to pseudobulges that tend to have smaller black holes and greater scatter than for classical bulges of the same stellar mass. Therefore, we caution that the inferred $M_\mathrm{BH}$ values may be overestimates if our sample contains pseudobulges or some pseudobulge components.

For normal single AGNs, the bolometric luminosity, \lbol, can usually be estimated from either optical and X-ray luminosities.
However, for our targets, the hard X-ray luminosities are much lower than for single AGNs with the same \OIII\ luminosities (paper I).
Three possible reasons are:
(1) the observed X-rays are low due to high gas absorption (e.g., \citealt{Bassani1999ApJS});
(2) the X-rays are intrinsically weak due to high accretion state that radiates little coronal emission (e.g., \citealt{Desroches2009ApJ, Dong2012ApJ});
(3) there are significant \OIII\ luminosity excesses due to shocks in our galaxy mergers (e.g., \citealt{Dopita1995ApJ}).
The emission line diagnostic ratios featuring \OIII/\hbeta\ and \NII/\halpha\ show that our targets are well in the regime of Seyferts, with only one displaying relatively large uncertainty \citep{Liu2010ApJL}.
It is unlikely that the emission lines of our AGNs are mainly powered by shocks, as they do not lie in the region of LINERs on the diagnostic diagrams \citep{Ho1993ApJ}.
It is more likely that our targets are highly obscured rather than in a high accretion state, since the estimated Eddington ratios (Table \ref{tab:bol}; see discussion below) are not close to unity.  Additionally, our targets are all detected by the \textit{Wide-field Infrared Survey Explorer} (\textit{WISE}; \citealt{Wright2010AJ}) with $\lambda L_\lambda (12 \,\mu m) \sim 10^{44} \mathrm{erg \, s^{-1}}$.
The X-ray-MIR ratios of our target binary AGNs ($\lesssim 0.01$) are much smaller than normal AGNs ($\sim 0.3$; \citealt{Horst2008AA}), indicating that the X-ray emission of the AGNs are absorbed and reprocessed into IR emission.
As shown by \citet{heckman05} at low redshift, selection by narrow optical emission lines will recover most AGNs selected by hard X-rays (with the exception of BL Lac objects). On the other hand, selection by hard X-rays misses a significant fraction of the local AGN population with strong emission lines.
In view of the uncertainties associated with X-ray absorption in gas-rich mergers, we estimate the bolometric luminosities of the binary AGNs from their extinction-corrected \OIII\ luminosity (paper I),
\begin{equation}\label{eq:cbol}
\frac{L_\mathrm{bol}}{10^{40} \, \mathrm{erg\,s}^{-1}} = 112
 \left(\frac{L^{\rm cor}_\text{[{\scriptsize O} {\tiny III}]}}{10^{40} \, \mathrm{erg\,s}^{-1}}\right)^{1.2},
\end{equation}
with intrinsic scatter $0.4 \, \mathrm{dex}$ (see \citealt{Trump2015arXiv} and references therein). The results are listed in Table \ref{tab:bol}.
We adopt extinction-corrected rather than observed \OIII\ luminosity (e.g., \citealt{Stern2012MNRAS}) to estimate the bolometric luminosity, since the former provides a smaller scatter.

\begin{figure}[!htpb]
\begin{center}
\includegraphics[width=0.45\textwidth]{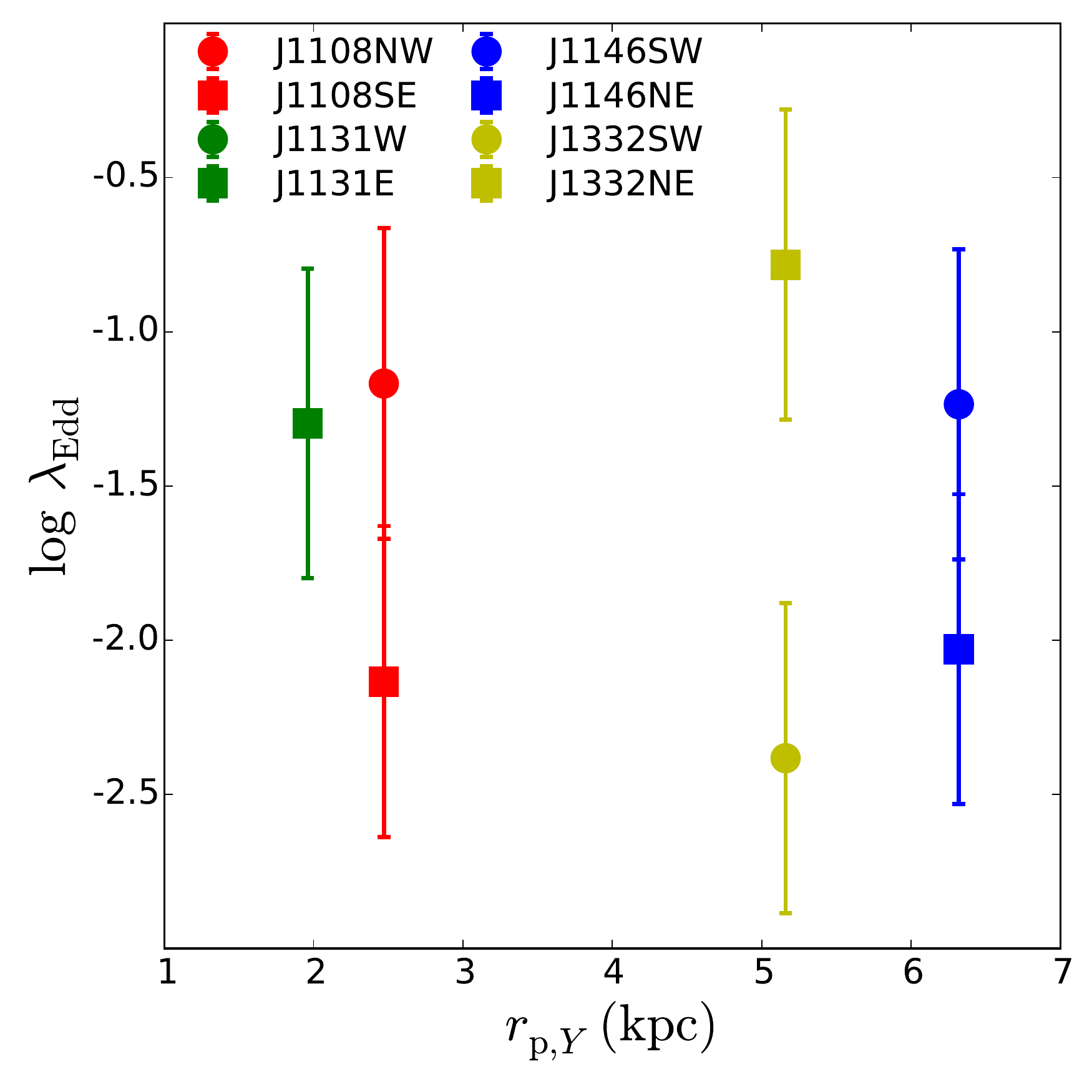}
\caption{
Eddington ratios of the double nuclei as a function of the nuclear separation measured in the $Y$ band (Table \ref{tab:obs}).
The Eddington ratios of all the nuclei are measured, except for J1131W (Table \ref{tab:bol}).
No apparent correlation is seen, which is not surprising given our small sample size and the limited dynamic range in nuclear separation.
}
\label{fig:bol1}
\end{center}
\end{figure}

We calculate the Eddington ratio, $\lambda_\mathrm{Edd}=L_{\rm bol}/L_{\rm Edd}$, for each nucleus (Table \ref{tab:bol}, Col. 8), with a typical uncertainty of $0.5 \, \mathrm{dex}$, the quadrature sum of the uncertainties of SMBH mass ($\sim 0.3 \, \mathrm{dex}$) and bolometric luminosity ($\sim 0.4 \, \mathrm{dex}$).
The Eddington ratios are plotted in Figure \ref{fig:bol1} against the nuclear separations. We find no significant correlation. A larger sample with a larger dynamic range in nuclear separation would be needed to reveal any correlation \citep{Ellison2011MNRAS,Liu2012}.
The Eddington ratios could be underestimated since the bulge stellar mass, and by extension, the SMBH masses, could be overestimated (\S \ref{subsec:mass}).
Nevertheless, the correlation between $\lambda_\mathrm{Edd}$ and \loiii\ of our binary AGNs, discussed below, indicates that our estimates are reasonable.

\begin{figure*}
\begin{center}
\includegraphics[width=0.95\textwidth]{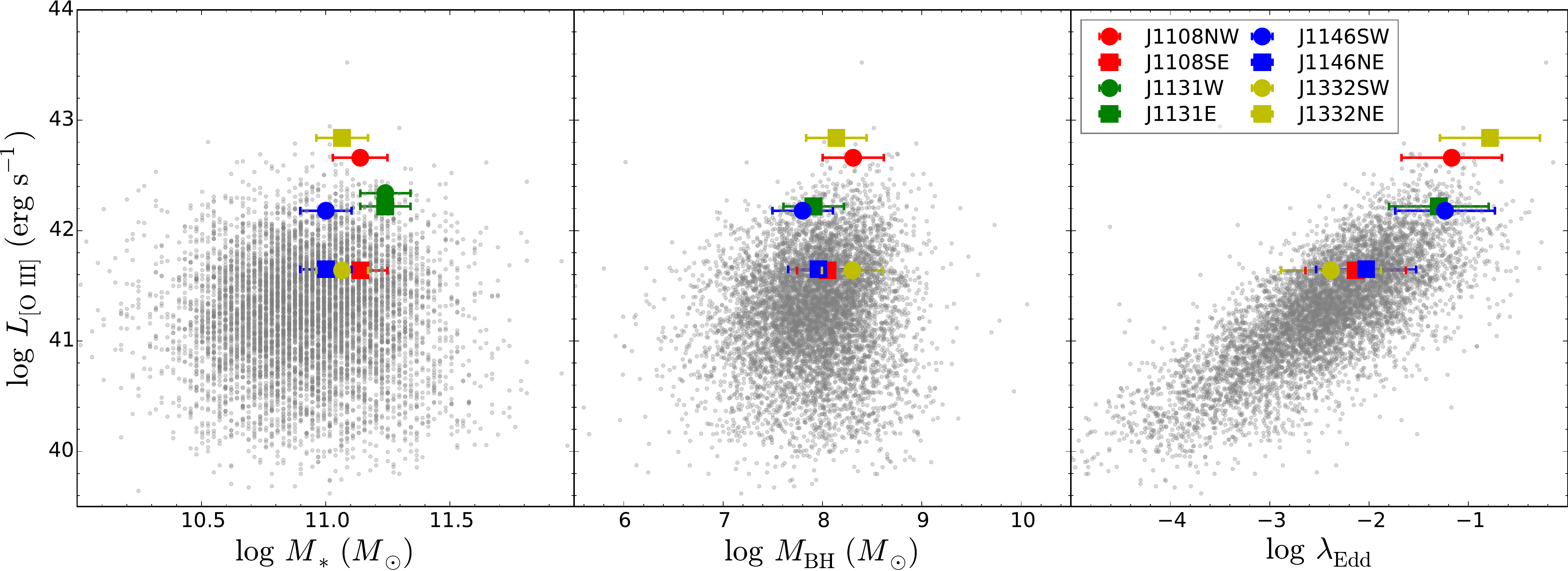}
\caption{
Comparison of our targets (large color symbols) with control samples of single AGNs from the MPA/JHU SDSS DR7 AGN sample (small grey points) within the same redshift range ($0.1<z<0.21$).
The left panel shows the galaxy stellar mass versus the extinction-corrected \OIIIb\ luminosity.
The middle and right panels show the SMBH masses and Eddington ratios against \loiii , but with the galaxy stellar masses within the same range as our targets ($10.5 < \mathrm{log}\,(M_*/M_\odot) < 11.5$).
}
\label{fig:bol2}
\end{center}
\end{figure*}

We compare our targets with control samples of single AGNs using the MPA/JHU SDSS DR7 AGN sample\footnote{\url{http://www.mpa-garching.mpg.de/SDSS/DR7/}} within the same redshift range ($0.1<z<0.21$).
Figure \ref{fig:bol2} shows that our host galaxies have typical stellar masses (\textit{left}) and SMBH masses estimated from stellar velocity dispersions (\textit{middle}).
While occupying the more luminous end in \loiii\ by selection \citep{Liu2010ApJ}, our targets follow the correlation of $\lambda_\mathrm{Edd}$ and \loiii\ for single AGNs (Figure \ref{fig:bol2} \textit{right}).
This may indicate that our SMBH mass and $\lambda_\mathrm{Edd}$ estimates are reasonable.
Our targets show moderate Eddington ratios, falling in the range of single AGNs.
While mergers are expected to promote gas inflow and trigger a higher rate of BH accretion than moderately luminous AGNs fueled by secular instability \citep{Hernquist1989Natur, Kauffmann2000MNRAS, Hopkins2008ApJS}, the tidal enhancement in accretion luminosity can be a subtle effect, in particular for mergers in their relatively early stages where the merging components are still separable \citep[e.g.,][]{Ellison2011MNRAS,Liu2012}.


\begin{table*}
\caption{Host Galaxy Bulge and Black Hole Accretion Properties of the Four kpc-scale Binary AGNs}
\begin{center}
\begin{tabular}{c c c c c c c c}
  \hline
  \hline
Target Name & log \loiiiOBS & log \loiiiCOR & log \lbol & $M_{z, \mathrm{bulge}}$& ${\rm log} ~ M_{\rm *, bulge}$ & ${\rm log} ~ M_{\rm BH}$ & \REdd \\
                   & \lum  & \lum  & \lum  &  (mag)  & ($M_\odot$) & ($M_\odot$) &         \\
(1)                & (2)   & (3)   & (4)   & (5)     & (6)         & (7)         & (8)     \\ \hline
J1108NW  & 42.16 & 42.66 & 45.24 & $-$21.96  & 10.67       & 8.31        & 0.068   \\
J1108SE  & 41.52 & 41.64 & 44.02 & $-$21.41  & 10.45       & 8.05        & 0.007   \\
J1131W & 41.40 & 42.34 & 44.86 & \nodata & \nodata     & \nodata     & \nodata \\
J1131E & 41.31 & 42.22 & 44.71 & $-$20.77  & 10.33       & 7.91        & 0.050   \\
J1146SW  & 41.93 & 42.18 & 44.67 & $-$20.77  & 10.24       & 7.80        & 0.058   \\
J1146NE  & 41.38 & 41.65 & 44.03 & $-$21.11  & 10.37       & 7.96        & 0.009   \\
J1332SW  & 41.06 & 41.64 & 44.02 & $-$21.86  & 10.67       & 8.30        & 0.004   \\
J1332NE  & 41.83 & 42.84 & 45.46 & $-$21.52  & 10.53       & 8.14        & 0.165   \\ \hline
\end{tabular}
\end{center}
\tablecomments{
Col. 2 \& 3: observed and extinction-corrected \OIIIb\ emission-line luminosity measured from spatially resolved optical spectra (Paper I).
Col. 4: AGN bolometric luminosity estimated from \loiiiCOR\ based on Equation \ref{eq:cbol}.
Col. 5: $z$-band absolute magnitudes of the bulge components for each galaxy obtained through \galfit\ decomposition (see Table \ref{tab:galfit}).
We have applied k-corrections to convert the original $Y$-band magnitudes measured by \galfit\ to the SDSS $z$ band.
For those nuclei decomposed with one PSF and one \sersic\ component, the latter is regarded as the bulge component.
For the Nucleus 2 of J1131 and J1146 in which two \sersic\ components are employed, we choose the inner component with higher \sersic\ index and smaller $R_\mathrm{e}$ as the bulge component.
Nucleus 1 of J1131 is fitted by a single PSF, so no bulge component is measured.
Col. 6: Bulge stellar masses estimated from $M_{z, \mathrm{bulge}}$, following the method of \cite{Bell2003ApJS} adopting the overall colors of the galaxies as a proxy for the colors of the bulges (Col. 7 in Table \ref{tab:mag}; see the text for more details).
Col. 7: SMBH mass estimated from $M_{*, bulge}$ based on the relation in \cite{Kormendy2013ARAA}.
Col. 8: Eddington ratio, $\lambda_\mathrm{Edd}=L_\mathrm{bol}/L_\mathrm{Edd}$.
}
\label{tab:bol}
\end{table*}

\section{Discussion}\label{sec:discuss}

\subsection{X-ray Emission and Host-galaxy Color}
\label{subsec:x+c}

Paper I has reported the X-ray emission by \chandra\ in both nuclei of J1108, both nuclei of J1146, and J1332NE.\footnote{Recently \cite{Bondi2016arXiv} have reported the presence of one radio AGN associated with J1131E based on Very Long Baseline Interferometry (VLBI) using the Eutropean VLBI Network at 5 GHz. They have also identified a possible compact radio core in J1108NW based on VLA observations at three frequencies. Since, given the accretion properties (e.g., $M_\mathrm{BH}$ and X-ray luminosity), the radio emission of an AGN scatters considerably (e.g., \citealt{Merloni2003MNRAS, Zakamska2016MNRAS}), the non-detection of radio AGN in some of the nuclei is not surprising.} 
Hard X-ray emission tends to coincide with nuclei displaying redder color (J1108SE, J1146SW and J1332NE), whereas soft X-rays are preferentially found in bluer nuclei (J1108NW and J1146NE).
As we discussed in \S \ref{subsec:mph}, the color in the nuclei is mainly determined by star formation and dust attenuation.
The soft X-ray excess in the bluer nuclei (J1108NW and J1146NE) indicates that nuclear star formation could contribute considerable amounts of soft X-rays.
Large amounts of gas may be associated with the observed nuclear star formation. Meanwhile, the X-ray radiation from the AGN could be heavily absorbed.
This may explain the deficit of hard X-ray emission in the bluer nuclei.
For the redder nuclei, the hard X-ray emission from AGNs implies that there is less gas column along the line of sight than in the bluer nuclei.
The redder colors are likely due to the relatively low star formation rates.
Dust attenuation may be an additional reason.
Meanwhile, if the blue clumps are mainly due to the AGN scattered light, the coincidence of soft and hard X-ray emission along with the UV emission is expected, which is not the case for our targets.
The Eddington ratios of our AGNs are normal, so that it is unlikely that the AGNs in our targets are intrinsically weak in the X-rays.


\subsection{Biases in the Samples of X-ray Selected AGNs}
\label{subsec:bias}

It has long been controversial whether there is a connection between mergers and moderate-luminosity AGNs.
One conclusion from the investigations of host galaxies of \chandra\ or \xmmnewton\ X-ray selected AGNs is that the host galaxies do not show statistically more disturbed features than do control samples \citep{Grogin2005ApJ, Gabor2009ApJ, Cisternas2011ApJ, Schawinski2011ApJ, Kocevski2012ApJ, Villforth2014MNRAS, Bruce2015arXiv}.
However, studies based on AGNs selected in other wavelengths, e.g., in the optical \citep{Ellison2008AJ, Ellison2011MNRAS, Ellison2013MNRAS}\footnote{\cite{Boehm2013AA} investigated optically selected AGNs and concluded that the morphologies of AGN hosts are different from those of interacting galaxies. However, their sample only consists of 21 type 1 AGNs.}
and radio \citep{Almeida2011MNRAS, Chiaberge2015ApJ}, seem to prefer a higher fraction of mergers in the host galaxies, although the difference may be subtle and requires a large sample to reveal.

There is growing evidence that binary AGNs are systematically weaker in the X-rays than single AGNs with comparable \loiii\ (\citealt{Teng2012ApJ}; paper I; \citealt{Comerford2015ApJ}).
This may be, on one hand, due to heavy absorption, because mergers may drive large amounts of gas into the nuclei.
On the other hand, large amounts of gas may fuel very strong accretion onto the SMBH and, therefore, the UV -- X-ray spectrum (from the {\it slim disk}; \citealt{Abramowicz1988ApJ}) should be much softer than the normal case ({\it standard disk}; \citealt{Shakura1973AA, Haardt1993ApJ, Shemmer2006ApJ, Desroches2009ApJ, Dong2012ApJ}).
For the same reason, we would expect that a large fraction of merger-triggered single AGNs should also be X-ray weak.
Therefore, one possible reason for the tension is that X-ray selected AGN samples may be biased against merger triggered AGNs.
In fact, most of the works use $L_\mathrm{hard \, X-ray} \gtrsim 10^{42} \mathrm{erg \, s^{-1}}$ to select their AGN samples.
When this criterion is applied to our binary AGN sample, only J1108 and J1146 are marginally qualified (paper I).
Based on a relatively small sample of 167 local galaxies, \cite{Koss2010ApJ, Koss2012ApJ} investigate the host galaxies of \swift\ BAT hard X-ray selected AGNs and find a larger fraction of mergers in AGN hosts than in normal galaxies.
The energy range of \swift\ BAT is $14\sim195$ keV \citep{Baumgartner2013ApJS}, much higher than \chandra\ and \xmmnewton.
The photons at that energy range are able to escape from gas absorption.
\cite{Kocevski2015arXiv} also find a higher merger fraction in the host galaxies of Compton thick AGN sample.
In addition, differences in the AGN and control samples, as well as in the morphological analysis methods, could affect the inferred merger rate \citep{Rosario2015AA}.
{\cite{Silverman2011ApJ} studied X-ray AGNs in a sample of kinematic galaxy pairs from the zCOSMOS and found a higher AGN fraction among galaxy pairs than control samples of single galaxies.
Since the galaxy pairs in their sample, by selection, have large projected separations ($\gtrsim 30 \, \mathrm{kpc}$; Fig. 5), the effect of tidal disturbance on gas is likely to be still small and therefore it is possible that AGNs in their sample are not in particular highly absorbed in the X-rays.
Finally, if the majority of merger-induced accretion luminosity only happens after the galaxies have merged as suggested by simulations \citep[e.g.,][]{Hopkins2008ApJS}, then we may be only witnessing the ``tip of the iceberg'' of the tidal enhancement by looking at host galaxies of binary AGNs where the merging components are still separable.
Further studies based on larger samples are needed to draw firm conclusions to resolve the controversial results concerning the causal link between mergers and moderate AGN activity.

\section{Summary}\label{sec:sum}

This is the second paper in a series in which we present \chandra\ and \hst\ imaging of four representative optically selected kpc-scale binary AGNs drawn from the sample of \cite{Liu2010ApJL}. In Paper I, we have examined the nature of the ionizing sources in the double nuclei. In this paper, we have investigated the host galaxy morphology and AGN properties. 
We summarize our main findings as follow.

\begin{itemize}
\item We have employed \galfit\ to decompose the $Y$-band images of the host galaxies to study their structural properties.
The $Y$-band images of all four galaxies show double stellar nuclei and tidal features consistent with mergers in the relatively early stages.
Based on $Y$-band luminosities and $U-Y$ colors, we have estimated stellar masses for different structural components of the galaxies, stellar mass ratios between the merging components, and stellar-mass bulge-to-total ratios.
Three of the four targets are major mergers; the other one is a minor merger with a dominant disk component.
Based on the estimated bulge-to-total stellar mass ratios, the three major mergers would likely become bulge dominated after they become relaxed.
In contrast to the prototypical examples of binary AGNs which either involve gas-rich mergers of disks \citep[e.g., NGC 6240;][]{Komossa2003ApJL} or ellipticals at the centers of galaxy clusters that power twin jets \citep[e.g., 3C 75][]{owen85}, at least three of the four binary AGNs reside in intermediate type host components as indicated by their \sersic\ indices (Table \ref{tab:galfit}).
While drawn from a unique sample of AGNs with double-peaked narrow emission lines, the host galaxies of at least three of our four targets have similar morphological properties to those of single AGNs at similar redshifts, i.e., intermediate between pure disks and ellipticals.

\item Based on the empirical correlation of bulge stellar mass and SMBH mass, we have estimated the mass of SMBHs residing in the resolvable stellar bulges.
Combined with AGN bolometric luminosities estimated from \OIIIb\ luminosities corrected for extinction, we have estimated the Eddington ratios for each nucleus.
Compared with a representative control sample drawn at the same redshift and stellar mass, the AGN luminosities and Eddington ratios of our binary AGNs are similar to single AGNs.
While by selection our targets are biased against minor mergers, the fact that one of the four targets is a minor merger suggests that enhanced accretion is more likely in the minor component \citep[see also][]{barth08}, with the caveat that our sample size is small.


\item The $U-Y$ color maps display disturbed structures and clumpy star forming regions, which may contribute to excess emission in the soft X-rays but absorb hard X-ray emission from the AGNs prominently. Our binary AGNs are X-ray weak compared with single AGNs with similar \OIIIb\ luminosity, probably because of this gas absorption.
Considering the same situation may also happen to merger triggered single AGNs, we suggest that samples of X-ray selected AGNs may suffer from biases against gas-rich mergers.
\end{itemize}

\acknowledgments

We thank an anonymous referee for helpful comments.
JS thanks Yang Huang, Yiqing Liu, Long Wang and Jiayi Sun for helpful discussions and suggestions.
LCH acknowledges support by the Chinese Academy of Science through grant No. XDB09030102 (Emergence of Cosmological Structures) from the Strategic Priority Research Program and by the National Natural Science Foundation of China through grant No. 11473002.

Support for this work was provided by NASA through Chandra Award Number GO1-12127X issued by the \chandra\ X-ray Observatory Center, which is operated by the Smithsonian Astrophysical Observatory for and on behalf of NASA under contract NAS 8-03060.
Support for program number GO 12363 was provided by NASA through a grant from the Space Telescope Science Institute, which is operated by the Association of Universities for Research in Astronomy, Inc., under NASA contract NAS 5-26555.

This research has made use of software provided by the Chandra X-ray Center in the application packages CIAO, ChIPS, and Sherpa.

Funding for the SDSS and SDSS-II has been provided by the Alfred P. Sloan Foundation, the Participating Institutions, the National Science Foundation, the U.S. Department of Energy, the National Aeronautics and Space Administration, the Japanese Monbukagakusho, the Max Planck Society, and the Higher Education Funding Council for England.
The SDSS Web Site is http://www.sdss.org/.


Facilities: CXO (ACIS), HST (WFC3), Sloan

\bibliography{biAGN}


\end{document}